\documentclass[cits]{PoS}
\usepackage{graphicx}
\def\degr{\hbox{$^\circ$}}
\def\fdg{\hbox{$.\!\!^\circ$}}
\def\arcsec{\hbox{$^{\prime\prime}$}}
\def\farcs{\hbox{$.\!\!^{\prime\prime}$}}
\title{Multi-frequency VSOP and VLBI observations \\ of the quasar
	3C\,309.1}

\ShortTitle{Observations of the quasar 3C\,309.1}

\author{\speaker{Marcin P. Gawro\'nski}\\
        Toru\'n Centre for Astronomy, Nicolaus Copernicus University, Poland\\
        E-mail: \email{motylek@astro.uni.torun.pl}}

\author{Andrzej J. Kus\\
        Toru\'n Centre for Astronomy, Nicolaus Copernicus University, Poland\\
        E-mail: \email{ajk@astro.uni.torun.pl}}

\abstract{We present multi-frequency observational results for the Compact Steep-Spectrum (CSS)
quasar 3C\,309.1. The observations were carried out with the VSOP
at 1.6 and 4.8\,GHz and the VLBA at 15.4\,GHz. The source has a  
distorted, one-sided radio jet. Relativistics effects and physical 
properties are discussed. Comparing the predicted and observed X-ray fluxes 
in the framework of the synchrotron self-Compton model we derive
the beaming factor for 3C\,309.1\,. The complex structure of the jet may 
be explained by a relativistic helical flow in a heterogeneous, clumpy ISM.}

\FullConference{The 8th European VLBI Network Symposium \\
                September 26-29, 2006\\
		Toru\'n, Poland}

\begin{document}

\section{Introduction}
The quasar 3C\,309.1 belongs to the class of Compact Steep-Spectrum (CSS)
radio sources, which includes objects with projected linear sizes $<20$\,kpc and steep radio  
spectra ($\alpha > 0.5$,$~S\propto \nu^{-\alpha}$) \cite{kap81,pewa82}. Their fraction is high -- up to 30\% depending 
on the selection frequency. CSS sources are identified with quasars, radio galaxies and Seyferts and 
the morphological separation between CSS quasars and galaxies -- similar to that observed in the case of 
larger sources -- has been found: radio galaxies have simple double radio structures, whereas quasars show either triple structures
with a strong central component or complex structures \cite{spen89,fan90}. 
\par
3C\,309.1 is one of the most luminous object in the class ($L_{178\,\mathrm{MHz}}=8\times10^{28}~\mathrm{W}\,\mathrm{Hz}^{-1}$)
and its radio structure has been studied with the VLA, MERLIN and the VLBI \cite{kus81,kus90,aro95,lud98}. The overall
extent of the source is $2\farcs2$ which translates to $80\,\mathrm{kpc}$\footnote{We assume a cosmology with 
$H_0=75\,\mathrm{km}\,\mathrm{s}^{-1}\,\mathrm{Mpc}^{-1}$ $\Omega_M=0.3$, $\Omega_{\Lambda}=0.7$} for the redshift $z=0.905$ \cite{bur69}.
This quasar appears as a triple source in the VLA and MERLIN images. There is a dominant central compact region, a curved,
one-sided, 1\,arcsecond-long jet emerging to the east with the hot-spot at the end and a low surface brightness component with 
extended north-south structure. 

\begin{figure}[b]
\hskip -0.2cm
\includegraphics[scale=1.25]{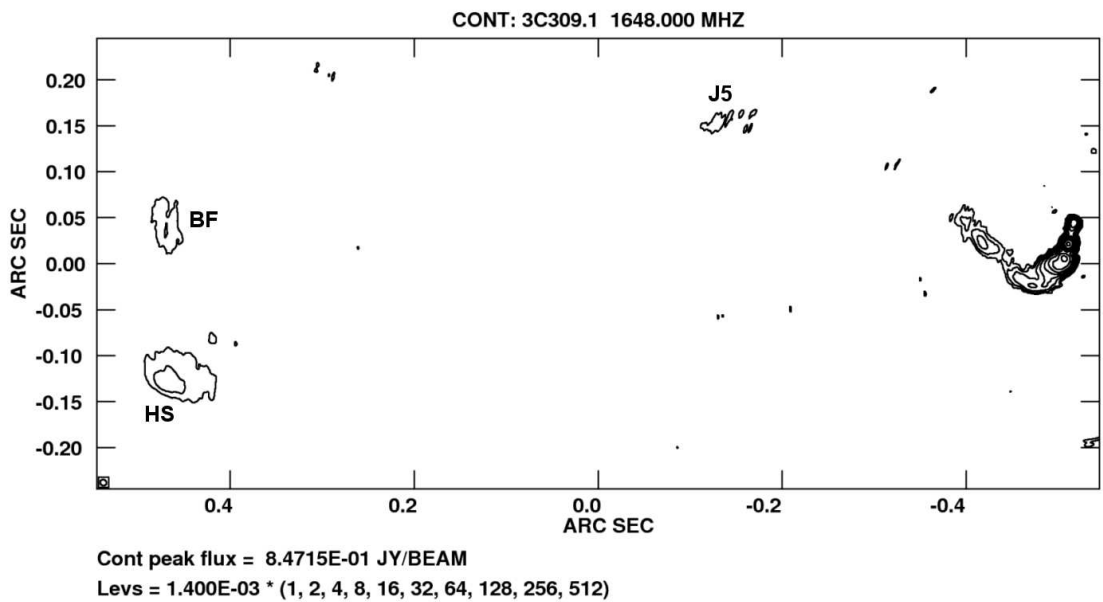}
\caption{Large-scale VLBI map of 3C\,309.1 at 1.6\,GHz.}
\label{l_band_large}
\end{figure}

\section{Observations}
\begin{figure}[t] 
\hskip .7cm
\includegraphics[scale=0.38]{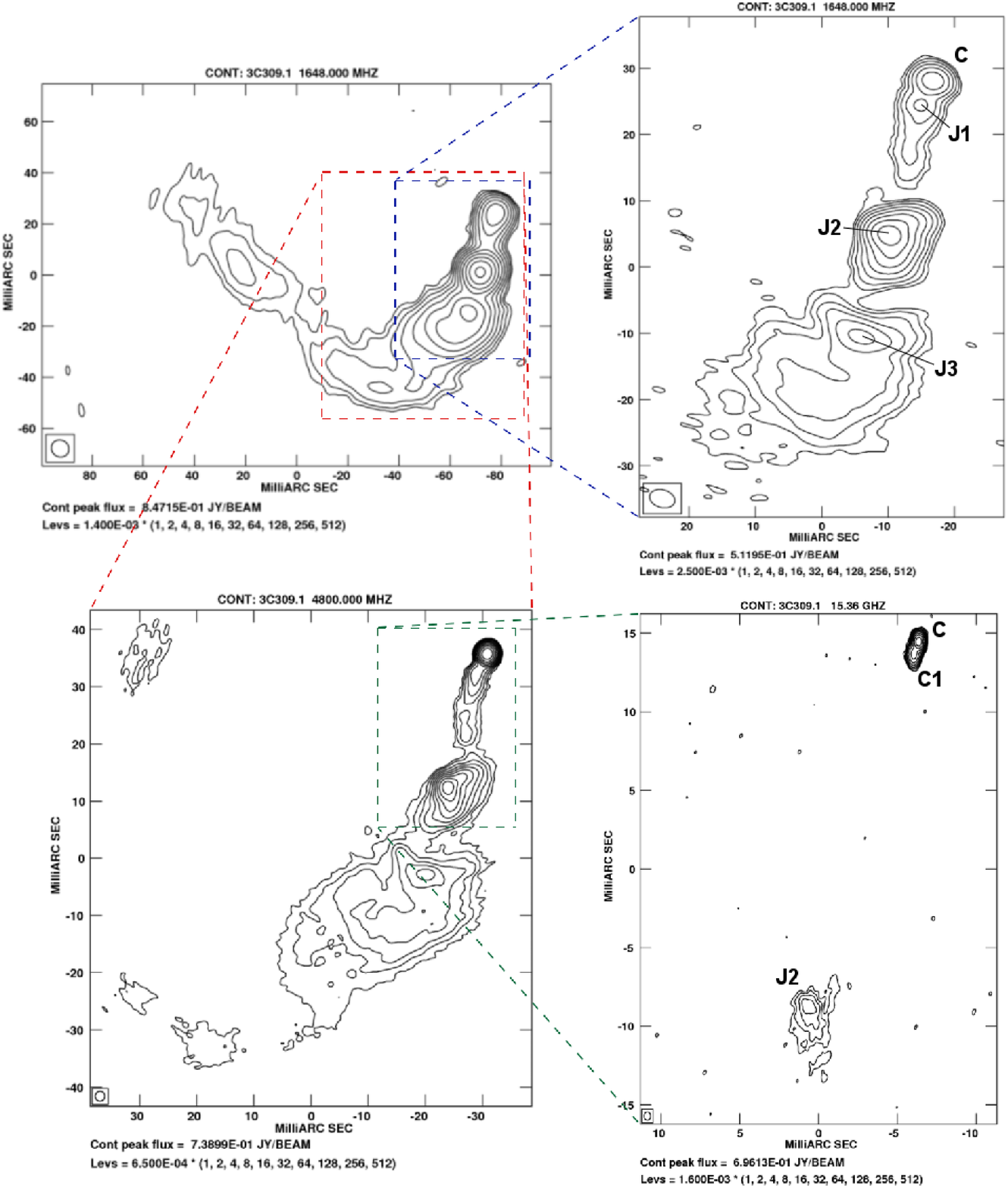}
\caption{The composition of VLBI and VSOP maps of 3C\,309.1. \emph{Upper left}: the inner part of the jet at 1.6\,GHz.
\emph{Upper right}: the VSOP map at 1.6\,GHz. \emph{Lower left}: the VLBI map at 4.8\,GHz. \emph{Lower right}: the
VLBI map at 15.4\,GHz. The main components are marked in the 1.6-GHz map. The superluminal feature C1 is marked
in the VLBI 15.4-GHz map.}
\label{complex_maps}
\end{figure}

We observed 3C\,309.1 with HALCA at 1.6\,GHz on 19 December 1997 and at 4.8\,GHz on 11 May 1998. 
We used all VLBA stations in both experiments with the addition of Goldstone 70\,m radio telescope at 1.6\,GHz and
two EVN stations (Effelsberg and Toru\'n) at 4.8\,GHz. The data from HALCA satellite were received by the tracking stations at
Goldstone, Tidbinbilla and Robledo. Clear fringes were found to HALCA at both epochs. We also carried out the VLBA observation
at 15.4\,GHz on 1 September 2002. The data were globally fringe fitted and calibrated in a standard way with AIPS. 
For the experiments including HALCA, we produced images for both: all the data and ground-based baselines only. The resulting images
of 3C\,309.1 are presented in Fig.\ref{l_band_large} and Fig.\ref{complex_maps}\,. We have labelled four main components 
of the jet as C (core), J1, J2 and J3.

\par
The ground-based map at 1.6\,GHz (Fig.\ref{l_band_large}) shows a complex jet structure and previously undetected 
features at some distance from the core C. The J5 component ($0\farcs4$ from the core at P.A$\simeq-75\degr$) is connected with 
the bend of the jet visible on the MERLIN maps \cite{lud98} whereas HS ($0\farcs98$  at  P.A$\simeq-90\degr$) and BF ($1\arcsec$ at 
P.A$\simeq-100\degr$) are probably a hot-spot and a backflow, respectively, similar to those observed in the standard FR\,II 
radio lobes. The inner part of the jet is dominated by components C and J2, which have similar brightness at 4.6\,GHz and 
are connected by a smoothly curving jet. There is a bright spot J1 of emission inside this part of the jet. Components J2 and J3 
are embedded in large regions of emission also detected in previous observations \cite{aro95}. The jet extends about 
80 mas at P.A $\simeq-155\degr$, rapidly changes its direction to P.A$\simeq-45\degr$ and extends another 
110 mas as seen on the ground-based VLBI map at 1.6\,GHz (Fig.\ref{l_band_large}).   

\begin{figure}[t] 
\hskip .5cm
\includegraphics[scale=1.0]{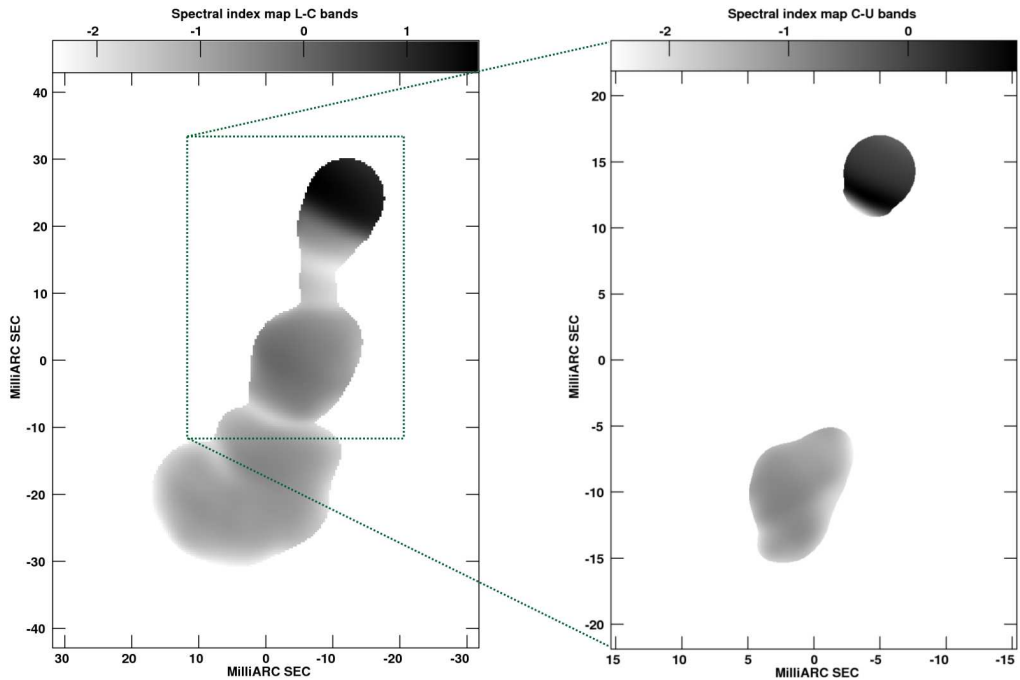}
\caption{The spectral index maps. \emph{Left}: between 1.6 and 4.8\,GHz. \emph{Right}: between 4.8 and 15.4\,GHz.}
\label{sp_in_maps}
\end{figure}

\section{Results}
The maps of the spectral index between 1.6 and 4.8\,GHz, and between 4.8 and 15.4\,GHz are shown in Fig.\ref{sp_in_maps}\,.
We noticed that the flattening of spectral index of the jet is connected with the increase of brightness. This could be an artifact 
introduced during the reduction process but a physical connection with the change of the jet direction is also possible. If we assume 
a stable helical path model \cite{kus90} this effect could easily be explained by the increased beaming 
factor and supports the idea of a stable jet path. Components J2 and J3 are those regions of the flow where the local velocity
vector is close to the line of sight. Theoretical studies of Kelvin-Helmholtz instabilities in the relativistic jets have
shown that they are able to create stable helical pattern \cite{har87}. It has been suggested \cite{sie04} that the origin of the 
infrared radiation is in the inner $\sim250\,\mathrm{pc}$, thus the helical part of the 3C\,309.1 jet is placed in a very dense medium. 
We discovered a relativistic motion of the new blob C1 by comparing our map obtained at 15.4\,GHz with the
MOJAVE project data \cite{kel04,lis05}, and found C1 apparent angular separation rate $\mu=0.15\pm0.02\,\mathrm{mas}/\mathrm{yr}$ 
which corresponds to $\beta_{app}=7.0\pm0.5\,c$\,. Such a behaviour was already suggested in \cite{kus90} and \cite{ros04}. No change in the position 
of component J2 relative to the core was detected.  
We derived the Doppler factor using the estimated synchrotron self-Compton X-ray flux. The self-Compton flux can be predicted 
based on the angular size of the components $\theta_r$, the synchrotron flux $F_r$ at frequency $\nu_r$, and the Doppler factor
 $\delta$. By comparing the observed and the predicted self-Compton fluxes one can derive $\delta$ in the case of moving sphere 
according to the formula \cite{ghi93}:
\begin{equation}
\delta=f(\alpha)\,F_r\,\biggl[\frac{ln(\nu_b/\nu_r)}{F_x\,\theta_r^{6+4\alpha}\,\nu_x^{\alpha}\,\nu_r^{5+3\alpha}}
\biggl]^{1/(4+2\alpha)}\,(1+z)
\end{equation}
where $F_x$ and $\nu_x$ are X-ray flux and frequency, $\nu_b$ is the synchrotron high frequency cut-off (assumed to be
$10^{14}$\,Hz), $\alpha$ is the spectral index of optically thin synchrotron emission and $f(\alpha)\simeq0.08\alpha+0.14$. Knowing 
the above, we estimated the Lorentz factor $\gamma$ which corresponds to the bulk velocity of the emitting plasma $\gamma$, and 
the angle $\phi$ between the line of sight and the direction of the bulk velocity ($\gamma=1.5$, $\phi=15\fdg5$). To evaluate these, we used the data  
from the Chandra archive, $F_{2-10\,\mathrm {keV}}=9.83\times\,10^{-8}\,\mathrm{Jy}$ obtained on 28 January 2002 \cite{shi05}. As the epochs of 
the Chandra and the VLBA observations at 15.4\,GHz were close, we choose these to calculate the ratios of the predicted to the observed 
X-ray flux for each main component. They are presented in Tab.\ref{tab} along with the brightness temperatures $T_b$ derived from 4.8\,GHz maps. 
If we assume that the orientation  of the central engine is similar to that of extended structures visible in the VLA and MERLIN maps then the overall 
size of 3C\,309.1 would be $\sim80\,\mathrm{kpc}$. Thus, it is plausible to assume that 3C\,309.1 is a small FR\,II radio galaxy oriented close to 
the line of sight and is similar to 3C\,216 \cite{bar88}.

\begin{table}
\begin{center}
\begin{tabular}{@{}c | l c@{}}
\hline
\hline
~~~~ Component ~~~~ & ~~~~ $T_B$~[K]~~(1)~~ & ~~~~~~~ $F_{th}/F_{X}$~~~~~~~~ \\
\hline
\hline
C & ~~~$2.6\times\,10^{11}$ & 10 \\
J1 & ~~~$7.6\times\,10^{8}$ & $<10^{-5}$ \\
J2 & ~~~$1.8\times\,10^{10}$ & $<10^{-5}$ \\
J3 & ~~~$7.9\times\,10^{8}$ & $<10^{-5}$ \\
\hline
\hline
\end{tabular}
\end{center}
\caption{The brightness temperature and the ratio of predicted to observed X-ray flux for main components.
Note: (1) We used maps at 4.8\,GHz.}
\label{tab}
\end{table}

\vskip .8cm
\par
We gratefully acknowledge the VSOP Project, which is led by the Institute of Space and Astronautical Science of
Japan Aerospace Exploration Agency, in cooperation with many organizations and radio telescopes around the world.
This work has made use of the VLBA, which is an instrument of the National Radio Astronomy Observatory, a facility
of the National Science Foundation, operated under cooperative agreement by Associated Universities, Inc. 
The European VLBI Network is a joint facility of European, Chinese and South African radio astronomy 
institutes funded by their national research councils. We are grateful to the group of the VLBA 2-cm Survey
and the group of the MOJAVE project for providing their data.


\begin{thebibliography}{99}

\bibitem{aro95} S.\,E.~Aaron,  J.\,F.\,C.~Wardle, D.\,H.~Roberts, Vistas Astron. {\bf41}, 225 (1997).

\bibitem{bar88} P.\,D.~Barthel, T.\,J.~Pearson, A.\,C.\,S.~Readhead, ApJ {\bf329}, 51 (1998).

\bibitem{bur69} G.\,R.~Burbidge, E.\,M.~ Burbidge, Nature {\bf222}, 735 (1969).

\bibitem{fan90} R.~Fanti, C.~Fanti, R.\,T.~Schilizzi et al., A\&A {\bf231}, 333 (1990).

\bibitem{ghi93} G.~Ghisellini, P.~Padovani, A.~Celotti, L.~Maraschi, ApJ {\bf407}, 65 (1993).

\bibitem{har87} P.\,E.~Hardee, ApJ {\bf318}, 78 (1987).

\bibitem{kap81}	V.\,K.~Kapahi, A\&AS {\bf43}, 381 (1981).

\bibitem{kel04} K.\,I.~Kellermann, M.\,L.~Lister, D.\,C.~Homan et al., ApJ {\bf609}, 539 (2004).

\bibitem{kus81} A.\,J.~ Kus, P.\,N.~Wilkinson, R.\,S.~Booth, MNRAS {\bf194}, 527 (1981).

\bibitem{kus90} A.\,J.~ Kus, P.\,N.~Wilkinson, T.\,J.~Pearson,, A.\,C.\,S.~Readhead in proceedings of \emph{Parsec-scale radio
jets}, p.161 (Cambridge University Press, Cambridge) (1990).

\bibitem{lis05}  M.\,L.~Lister, D.\,C.~Homan, AJ {\bf130}, 1389 (2005). 

\bibitem{lud98} E.~L\"udke, S.\,T.~Garrington, R.\,E.~Spencer et al., MNRAS {\bf299}, 467 (1998).

\bibitem{pewa82} J.\,A.~Peacock, J.\,V.~Wall, MNRAS {\bf198}, 843 (1982).

\bibitem{ros04} E.~Ros, \emph{Extending and exploring the 2cm Survey sample} in \emph{Proceedings of the 7th European VLBI
Network Symposium}(2004).

\bibitem{sie04} R.~Siebenmorgen, W.~Freudling, E.~Kr\"ugel, M.~Haas, A\&A {\bf421}, 129 (2004).

\bibitem{shi05} Y.~Shi , G.\,H.~Rieke , D.\,C.~Hines et al., ApJ {\bf629}, 88 (2005).

\bibitem{spen89} R.\,E.~Spencer, J.\,C.~McDowell, M.~Charlesworth et al., MNRAS {\bf240}, 657 (1989).

\end{thebibliography}
\end{document}